\documentclass[twocolumn,showpacs,showkeys,preprintnumbers,
amsmath,amssymb,aps,floatfix,jcp,reprint]{revtex4}
 
\usepackage{graphicx}
\usepackage{amsmath}
\usepackage{multirow}
\usepackage{times}
\usepackage{amssymb}
\usepackage{dcolumn}%
\usepackage{bm}%
\usepackage{hyperref}
\usepackage{ctable}
\usepackage{tabularx}

\hypersetup
{
	colorlinks=true,
	linkcolor=blue,
	citecolor=blue
}

\newcommand{\Cu}{$\mbox{Cu}$ }
\newcommand{\Zr}{$\mbox{Zr}$ }
\newcommand{\CuZr}{$\mbox{CuZr}$ }

\begin{document}

\title[]{{Fast dynamics perspective on the breakdown of the Stokes-Einstein law in fragile glassformers}}
\author{F.\@ Puosi  }
\affiliation{Univ. Grenoble Alpes, CNRS, Grenoble INP \footnote{Institute of Engineering Univ. Grenoble Alpes}, SIMaP, F-38000 Grenoble, France}
\author{A.\@ Pasturel  }
\affiliation{Univ. Grenoble Alpes, CNRS, Grenoble INP \footnote{Institute of Engineering Univ. Grenoble Alpes}, SIMaP, F-38000 Grenoble, France}
\author{N.\@ Jakse  }
\affiliation{Univ. Grenoble Alpes, CNRS, Grenoble INP \footnote{Institute of Engineering Univ. Grenoble Alpes}, SIMaP, F-38000 Grenoble, France}
\author{D.\@ Leporini}
\affiliation{Dipartimento di Fisica ``Enrico Fermi'', 
Universit\`a di Pisa, Largo B.\@Pontecorvo 3, I-56127 Pisa, Italy}
\affiliation{IPCF-CNR, UOS Pisa, Italy}

\date{\today}

\begin{abstract}

The breakdown of the Stokes-Einstein (SE) law in fragile glassformers is examined by Molecular-Dynamics simulations of atomic liquids and polymers and consideration of the experimental data concerning the archetypical OTP glassformer. All the four systems comply with the universal scaling between the viscosity (or the structural relaxation) and the Debye-Waller factor $\langle u^2\rangle$, the mean square amplitude of the particle rattling in the cage formed by the surrounding  neighbors. It is found that the SE breakdown is scaled in a master curve by a reduced $\langle u^2\rangle$.  Two approximated expressions of the latter, with no and one adjustable parameters respectively, are derived.

\end{abstract}

\maketitle

\section{Introduction}
\label{intro}

 {Under hydrodynamic conditions the diffusion coefficient $D$ is inversely proportional to the shear viscosity $\eta$. More quantitatively, the Stokes-Einstein (SE) relation states that the quantity  $D\eta/k_BT$ is a constant of the order of the size of the diffusing particle, $k_B$ being the Boltzmann constant \cite{Harris}}. Remarkably, despite its macroscopic derivation, SE accounts also well for the self-diffusion of many monoatomic and molecular liquids, provided the viscosity is low ( $\lesssim 10 \,Pa \cdot s$)  \cite{HansenMcDonaldIIIEd}. Distinctly, a common feature of several fragile glass formers is  the breakdown of SE for increasing viscosity, that manifests as a partial decoupling between the diffusion and viscosity itself \cite{Ediger00,CristianoSE,LadJCP12,Puosi12SE}. The decoupling is well accounted for by the fractional SE (FSE)  $D\sim \eta^{-\kappa} $ \cite{SillescuSEJNCS94} where the non-universal exponent $\kappa$ falls in the range $[0.5-1]$ \cite{DouglasLepoJNCS98}. 
The usual interpretation of the SE breakdown relies on dynamic heterogeneity (DH), the spatial distribution of the characteristic relaxation times $\tau$ developing close to the glass transition (GT) \cite{Ediger00,SillescuSEJNCS94,BerthieRev}. In metallic liquids it has been shown that the crossover from SE to FSE is coincident with the emergence of DHs \cite{LadJCP12,Hu_JAP2016}. 

The SE  law deals with long-time transport properties. Yet, several experimental and numerical studies evidenced universal correlations between the long-time relaxation and the fast (picosecond) dynamics as sensed by Debye-Waller (DW) factor $\langle u^2\rangle$, the collective \cite{PuosiLepoJCPCor12,PuosiLepoJCPCor12_Erratum} rattling amplitude of the particle within the cage of the first neighbours  \cite{HallWoly87,OurNatPhys,UnivPhilMag11,OttochianLepoJNCS11,Puosi12SE,DouglasCiceroneSoftMatter12,SpecialIssueJCP13,CommentSoftMatter13, Puosi_JCP17}.
 In particular, correlations are found in polymers \cite{OurNatPhys,lepoJCP09,Puosi11}, binary atomic mixtures \cite{lepoJCP09,SpecialIssueJCP13}, colloidal gels \cite{UnivSoftMatter11}, antiplasticized polymers \cite{DouglasCiceroneSoftMatter12} and water-like models \cite{Merabia_JCP17,Vogel_JCP17}. Strictly related correlation between long-time relaxation and  the shear elasticity are known \cite{Dyre06,Puosi12,ElasticoEPJE15,BerniniJPCM17Tvgamma}.
Building on these ideas, using Molecular-Dynamics (MD) simulations of a polymer model, some of us showed that the SE breakdown is well signaled by the DW factor $\langle u^2\rangle$ \cite{Puosi12SE}. Further, Douglas and coworkers demonstrated that it is possible to estimate the self-diffusion coefficient from linking the DW factor to the relaxation time and assuming that a FSE relation holds \cite{DouglasLocalMod16}.  In the same spirit, we also mention the method for estimating from  $\langle u^2\rangle$ data the characteristic temperatures of glass-forming  liquids, including that of the SE breakdown and the onset of DHs \cite{Zhang_JCP15,Puosi_CuZr}.

The present paper provides novel evidence of the vibrational scaling of the breakdown of SE law in terms of the DW factor $\langle u^2\rangle$ by combining MD simulations of atomic and polymeric fragile glassformers and experimental data of the archetypical glassformer OTP \cite{tolleRepProg01,Sillescu_JPCB97}.

The paper is organized as follows. In Sec. \ref{methods} details about the numerical models and the quantities of interest are given. The results are presented and discussed in Sec. \ref{results}.

\section{Models and methods}
\label{methods}
MD simulations for a Lennard-Jones binary mixture (BM) and the \CuZr metallic alloy (MA) were carried out using LAMMPS molecular dynamics software \cite{lammps}. As to BM, we consider a generic three-dimensional model of glass-forming liquid, consisting of a mixture of A and B particles, with $N_A=1600$ and $N_B=400$, interacting via a Lennard-Jones potential 
  $ V_{\alpha\beta}(r)= 4 \epsilon_{\alpha \beta}\left[ \left( \frac{\sigma_{\alpha\beta}}{r} \right)^{12} - \left( \frac{\sigma_{\alpha\beta}}{r} \right)^{6}  \right ] $
with $\alpha,\beta=A,B$ and $r$ being the distance between two particles.  The parameters $\epsilon_{AA}$, $\sigma_{AA}$ and $m_A$ define the units of energy, length and mass; the unit of time is given by $\tau_0=\sigma_{AA}\sqrt{(m_A/\epsilon_{AA})}$. We set $\epsilon_{AA}=1.0 $, $\epsilon_{AB}=1.5 $, $\epsilon_{BB}=0.5 $, $\sigma_{AA}=1.0$, $\sigma_{AB}=0.8$ and $\sigma_{BB}=0.88$ and $m_A=m_B=1$. It is known that, with this choice, the system is stable against crystallization \cite{KobAndersenPRE1}. The potential is truncated at $r=r_c=2.5$ for computational convenience. The total density $\rho=1.204$ is fixed and periodic boundary conditions are used. The system is equilibrated in the NVT ensemble and the production runs are carried out in the NVE ensemble.
As to MA, an embedded-atom model (EAM) potential was used to describe the interatomic interactions in the \CuZr binary alloy  \cite{MendelevJAP09}.  Each simulation consists of a total number of $23328$ atoms contained in a box with periodic boundary conditions. The initial configurations were equilibrated at $2000 \,\mbox{K}$ for $5\,\mbox{ns}$ followed by a rapid quench to $500 \,\mbox{K}$ at a rate of $10^{11}\,\mbox{K/s}$. The quench was performed in the $\mbox{NPT}$ ensemble at zero pressure. During the quench run configurations at the temperatures of interest were collected and, after adequate relaxation, used as starting points for the production runs in the $\mbox{NVT}$ ensemble.  

We consider the mean square particle displacement (MSD) $\Delta r^2(t)$ and define the Debye-Waller (DW) factor $\langle u^2\rangle=\Delta r^2(t_{DW})$ where $t_{DW}$ is a measure of the trapping time of a particle in the cage of the surrounding ones and equals the time at which $\log$ MSD vs $\log t$ has minimum slope \cite{OurNatPhys,SpecialIssueJCP13}. For the BM systems $t_{DW}\approx 1$ whereas for the MA system $t_{DW}\approx 1\,\mbox{ps}$, which is typical of metallic liquids.
The self-diffusion coefficient $D$ is determined  via the long-time limit $D=\lim_{t\rightarrow\infty}\Delta r^2(t)/6t$.  We define the structural relaxation time $\tau_\alpha$ via the relation $F_s(q_{max},t)=1/e$ where $q_{max}$ is the maximum of the static structure factor and $F_s$ the self part of the intermediate scattering function (ISF) \cite{OurNatPhys,SpecialIssueJCP13}.   
The degree to which particle displacements deviate from a Gaussian distribution is quantified by the non-gaussian parameter (NGP) $\alpha_2(t)=3 \Delta r^4(t) / 5 \Delta r^2(t)^2-1$ where $\Delta r^4(t) $ is the mean quartic displacement \cite{OurNatPhys}. The viscosity $\eta$ is calculated by integrating the stress autocorrelation function according to Green-Kubo formalism \cite{ZwanzingARPC65}, i.e. $\eta=(V/k_BT)\int_0^{\infty} \langle P_{\alpha\beta}(t_0)P_{\alpha\beta}(t_0+t)\rangle dt$ where $V$ is the volume, $P_{\alpha\beta}$ is the off-diagonal $\alpha\beta$ component of the stress and an average over the three components $\alpha\beta=xy,xz,yz$ is performed.

\begin{figure}[t]
\begin{center}
\includegraphics[width=0.95\columnwidth]{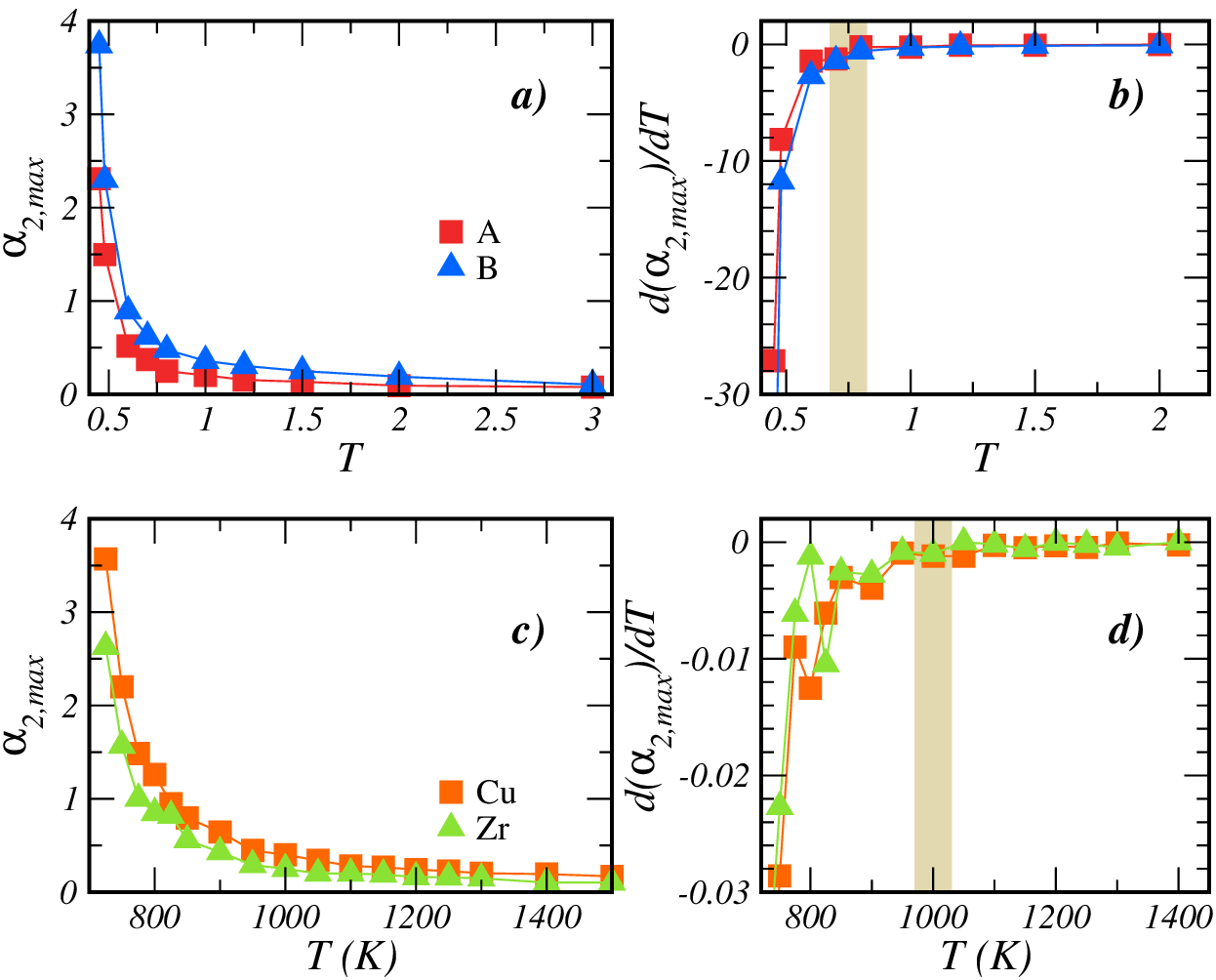}
\end{center}
\caption{   Panels a) and c):  Temperature dependence of $\alpha_{2,max}$, the maximum of the NGP, for the BM (a) and MA (c) systems. Panels b) and d):  temperature derivative $d\alpha_{2,max}/dT$ as a function of temperature for the BM (b) and MA (d) systems. The shaded regions mark the onset of dynamical heterogeneities at $T_{s}=0.75(5)$ for the BM and $T_{s}=1000(50)\,\mbox{K}$ for the MA, with no dependence on the species within our precision.}
\label{figDH}
\end{figure}

\section{Results and discussion}
\label{results}

First, we focus on the increase of DHs upon cooling as quantified by the NGP $\alpha_2$. The NGP time dependence has non-monotonous behavior: first it increases with time and then decays to zero in the gaussian diffusive regime, resulting in a maximum  $\alpha_{2,max}$ for times comparable to the structural relaxation time $\tau_\alpha$ \cite{OurNatPhys}. In Fig. \ref{figDH} (a,c) we plot the temperature dependence of $\alpha_{2,max}$ for  BM  and MA systems. Data are shown separately for each component of the two systems , A and B for BM and \Cu and \Zr for MA. The increase of $\alpha_{2,max}$ is slow at high temperature and accelerates as deeper supercooling is achieved. The crossover temperature $T_s$ can be detected from the temperature derivative $d\alpha_{2,max}/dT$, which is shown in Fig. \ref{figDH} (b,d) \cite{Hu_JAP2016}.  We find $T_{s}=0.75(5)$ for the BM and $T_{s}=1000(50)\,\mbox{K}$ for the MA, with no dependence on the species within our precision.
 
\begin{figure}[t]
\begin{center}
\includegraphics[width=0.95\columnwidth]{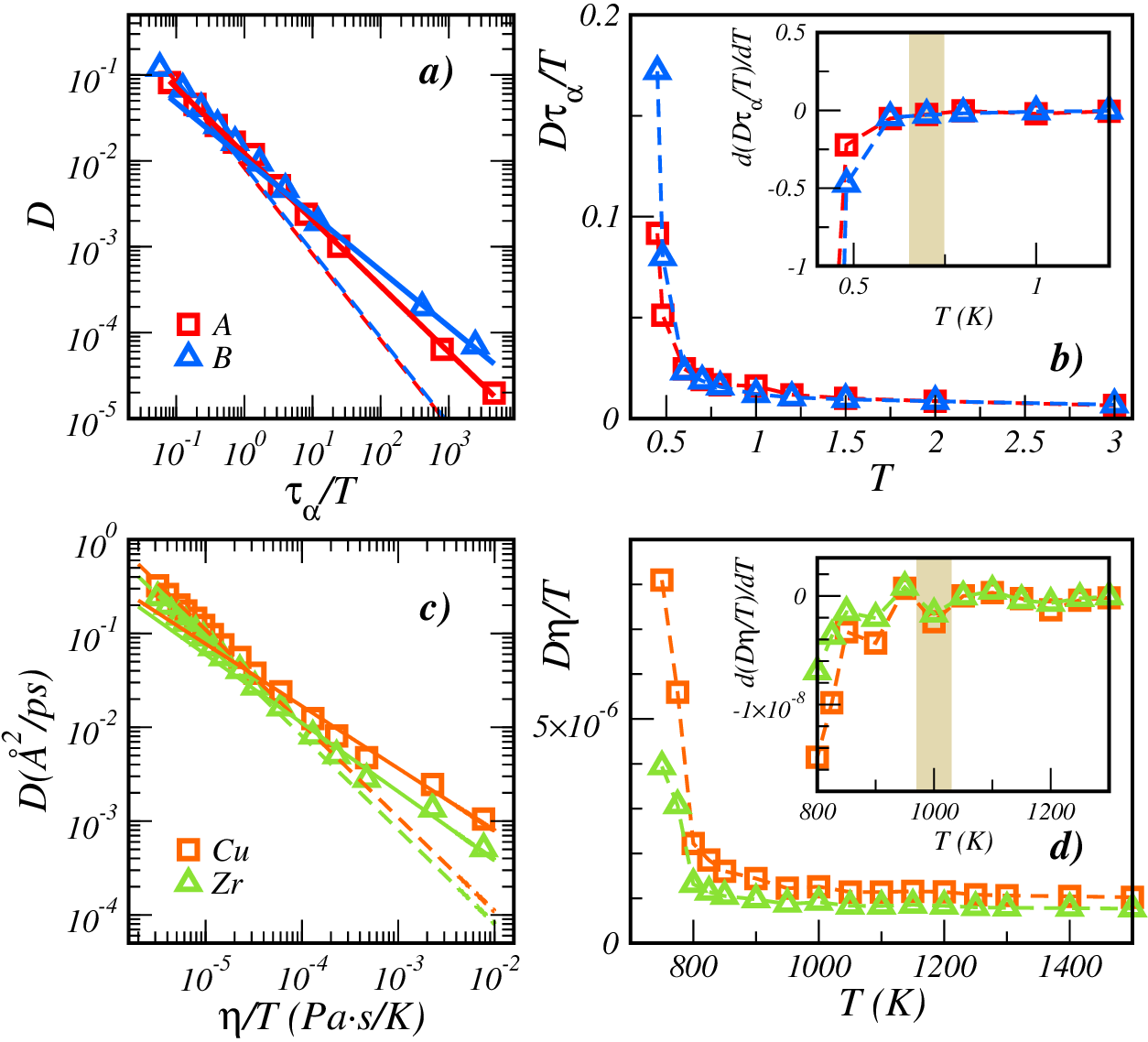}
\end{center}
\caption{ Panels a) and c):  Self-diffusion coefficient as a function of the ratio  $\tau_\alpha$/T (a) or $\eta/T$  (b) for the BM (a) and MA (c) systems. Dashed lines correspond to SE relation $D\propto (\tau_\alpha/T)^{-1}$ or $D\propto (\eta/T)^{-1}$. Full lines correspond to fractional SE relation $D\propto (\tau_\alpha/T)^{-\kappa}$ or $D\propto (\eta/T)^{-\kappa}$. Panels b) and d):  SE product $D \tau_\alpha/T$ (b) and $D \eta/T$ (d) versus temperature. Insets: corresponding temperature derivative $d(D \tau_\alpha/T)/dT$  and $d(D \eta/T)/dT$. Shaded regions mark the onset of the breakdown of SE relation.}
\label{figSE}
\end{figure}

Figure \ref{figSE} shows the decoupling of diffusion and viscosity in BM and MA. In both models, for each component,  the SE relation is obeyed at high temperature and breaks down in the supercooled regime. The decoupling is marked by a crossover towards a FSE relation $D\propto (\tau_\alpha/T)^{-\kappa}$ with $\kappa$ equal $0.77$ and $0.65$ for A and B particles respectively in the BM model and  $\kappa$ equal $0.66$ and $0.73$ for \Cu and \Zr atoms respectively in the MA model. It is worth noting that consideration of the ratio $\eta/T$ or $\eta$ alone in FSE is just a matter of convenience, given the huge change of viscosity in the small temperature range where FSE is observed.
The above results concerning the characteristic exponent $\kappa$ are intermediate between the prediction of the ``obstruction model'' $\kappa=2/3$  \cite{DouglasLepoJNCS98} and the universal value $\kappa=0.85$ found by Mallamace et al \cite{MallamaceFSEPNAS10}. The SE product $D \tau_\alpha/T$ and its temperature derivative $d(D \tau_\alpha/T)/dT$ reveals that the breakdown becomes apparent below $0.7$     and $1000\,\mbox{K}$ for the BM and MA models respectively. In both cases, this breakdown corresponds to the crossover temperature $T_s$ of the onset of DHs. 

\begin{figure}[t]
\begin{center}
\includegraphics[width=0.95\columnwidth]{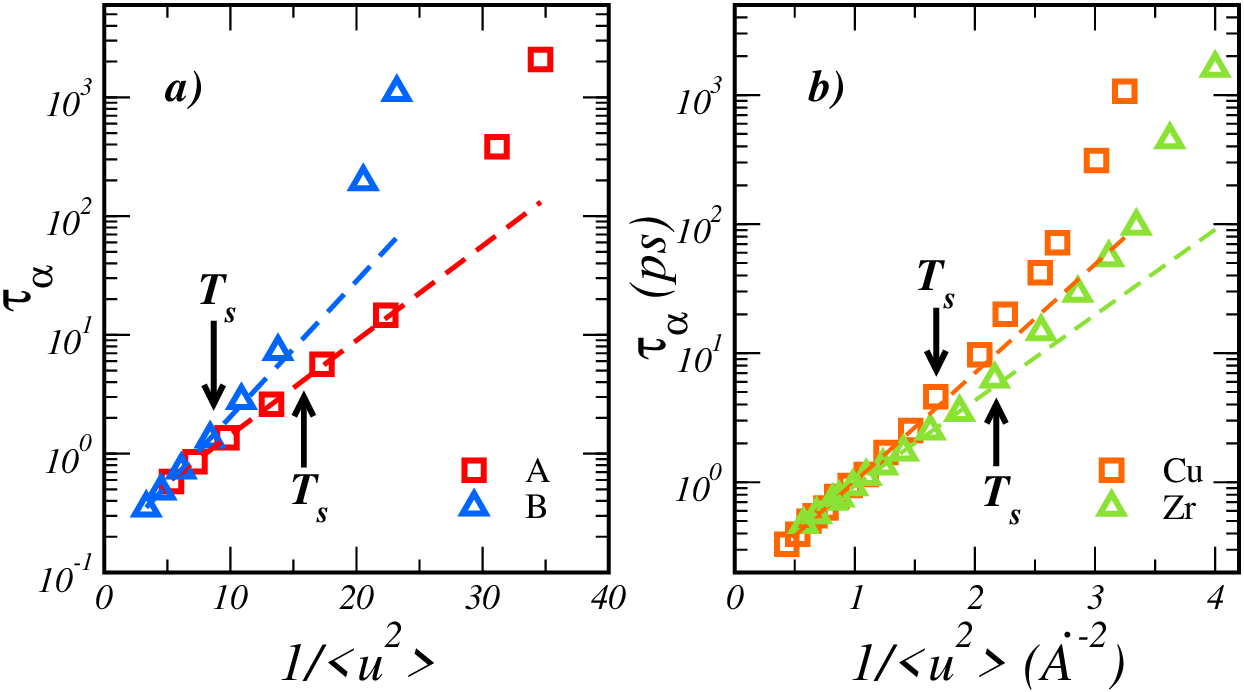}
\end{center}
\caption{ Panels a) and b): Structural relaxation time  as a function of the inverse DW factor  for the BM (a) and MA (b) systems. Dashed lines correspond to Eq. \ref{eqn:hw}. The location of the onset temperature $T_s$ is indicated for all the species.  }
\label{figUniv}
\end{figure}

%
%

Hall and Wolynes \cite{HallWoly87} first elaborated a vibrational model relating the slowing down on approaching GT with the accompanying decrease of the DW factor $\langle u^2 \rangle$ due to the stronger trapping effects \cite{HallWoly87}. They identified $\tau_\alpha$ with $\tau_\alpha^{(HW)}$ where:
\begin{equation}\label{eqn:hw}
\tau_\alpha^{(HW)}=\tau_0' \exp \left (  \frac{a^2}{2\langle u^2 \rangle}\right) 
\end{equation}
with $\tau_0'$ and $a^2$ adjustable constants. In particular, $a$ is the displacement to overcome the barrier activating the structural relaxation. We test Eq.\ref{eqn:hw} in Figure \ref{figUniv}.  For both BM and MA models, we find good agreement with simulation data if mobility is high (high $\langle u^2 \rangle$ or low $\tau_\alpha$). Otherwise, deviations become apparent, as already reported \cite{OurNatPhys,DouglasCiceroneSoftMatter12,SpecialIssueJCP13}. 
In particular, deviations from Eq. \ref{eqn:hw} correlate to the emergence of DHs in polymer melts  \cite{Puosi12SE}. This conclusion is in close agreement  with the finding that deviations from Eq. \ref{eqn:hw} become evident for both BM and MA models around the crossover temperature  $T_s$, see Figure \ref{figUniv}.

\begin{figure}[t]
\begin{center}
\includegraphics[width=0.9\columnwidth]{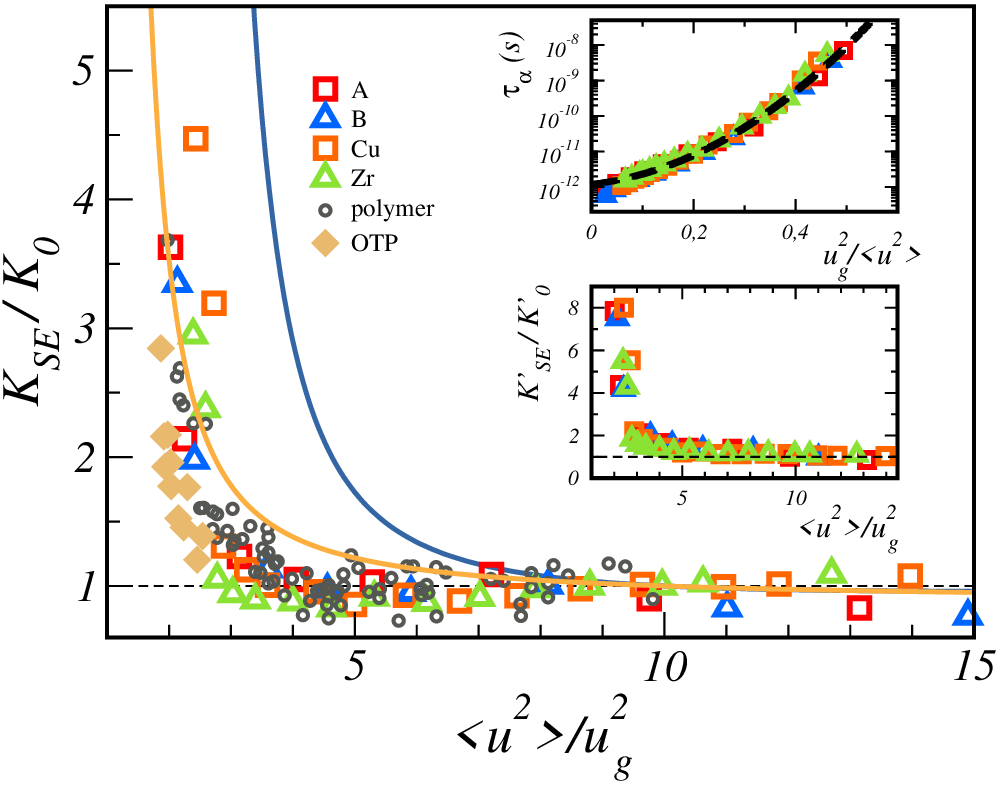}
\end{center}
\caption{ Main panel: Stokes-Einstein product $K_{SE}=D \eta$ (or $D \tau_\alpha$), normalized by its high temperature value $K_0$ ( $\tau_\alpha \simeq 1$ ps), as a function of the reduced DW factor  $\langle u^2 \rangle/u^2_g$, $ u^2_g$ being the DW factor at GT. In addition to the BM and MA systems, the plot shows numerical results concerning a model polymer melt \cite{Puosi12SE} and experimental data for ortho-terphenyl (OTP) \cite{tolleRepProg01,Sillescu_JPCB97}. Two predictions of the master curve are presented, namely Eq.\ref{eqn:Dtau} drawn by the the vibrational model of ref.\cite{OurNatPhys} with no adjustable parameter (dark-blue curve) and the FSE form $(\tau_\alpha/\tau_0) ^{1-\kappa}$ with $\kappa = 0.85$ (orange  curve). See text for details. Top inset: universal scaling between the  structural relaxation and the DW factor for the BM and MA systems ( the dashed line is Eq.\ref{eqn:parabola2} ). $ u^2_g$ is obtained by extrapolating Eq. \ref{eqn:parabola} to $T_g$ as done in \cite{OurNatPhys}. Bottom inset: alternative definition of the Stokes-Einstein product $K'_{SE}=D \eta/T$ (or $D \tau_\alpha/T$), normalized by its high temperature value $K'_0$, as a function of the reduced DW factor. 
}
\label{figSEUniv}
\end{figure}

An extension of Eq.\ref{eqn:hw} interprets the observed concavity of the curve $\log \tau_\alpha$ vs $1/\langle u^2 \rangle$ in Figure \ref{figUniv} as due the dispersion of the  $a$ parameter, modelled by a truncated gaussian distribution $p(a^2)$ with characteristic parameters $\overline{a^2}$ and $\sigma^2_{a^2}$ \cite{OurNatPhys,lepoJCP09,SpecialIssueJCP13}. Here, we define $\langle X \rangle_{a^2}$ the average of $X$ according to $p(a^2)$ and $\tau_\alpha = \langle \tau_\alpha^{(HW)} \rangle_{a^2}$. According to that approach, the  relation between $\tau_\alpha$ and the DW factor reads \cite{OurNatPhys,lepoJCP09,SpecialIssueJCP13}:
\begin{eqnarray}
\tau_\alpha&=&\tau_0\exp \left (  \frac{\overline{a^2}}{2\langle u^2 \rangle} + \frac{\sigma^2_{a^2}}{8\langle u^2 \rangle^2} \right) \label{eqn:parabola}\\
&=&\tau_0 \exp\left [ \hat{\beta} (u^2_g/ \langle u^2 \rangle) +\hat{\gamma} (u^2_g/ \langle u^2 \rangle)^2 \right ] \label{eqn:parabola2}
\end{eqnarray}
In Eq.\ref{eqn:parabola} $\tau_0$, $\overline{a^2}$ and $\sigma^2_{a^2}$ are system-dependent parameters. Eq.\ref{eqn:parabola} is recast in the {\it  universal} form given by Eq.\ref{eqn:parabola2} where $u^2_g$ is the DW factor at GT (defined via $\tau_\alpha = 10^2\, \mbox{s}$ or $\eta = 10^{12} \, \mbox{Pa}\cdot\mbox{s}$) \cite{OurNatPhys}. In particular, now the universal constants $\hat{\beta}=\tilde{\beta}\ln10=3.7(1)$ and $\hat{\gamma}=\tilde{\gamma}\ln10=28.4(2)$ are introduced, with $\tilde{\beta}$ and $\tilde{\gamma}$ defined in  \cite{OurNatPhys}, and $\tau_0$  ensures $\tau_\alpha = 10^2\, \mbox{s}$ at GT \cite{OurNatPhys}. 
Indeed, Eq. \ref{eqn:parabola2} was shown to provide a good description of experimental data in several systems \cite{OurNatPhys,lepoJCP09,SpecialIssueJCP13}. 

Now, we analyze the correlation between the SE breakdown and the fast dynamics. To this aim, we consider  the ratio $K_{SE}/K_0$ between  $K_{SE}=D \eta$ (or $K_{SE}=D \tau_\alpha$  when viscosity data are missing) and $K_0$, the quantity $K_{SE}$ evaluated at high temperature ( $\tau_\alpha \simeq 1$ ps). In Figure \ref{figSEUniv} we plot $K_{SE}/K_0$ 
 as a function of  $\langle u^2 \rangle/u^2_g$. We complement the MD results concerning the BM and MA models with literature data for few archetypical systems, specifically MD simulations of a model polymer melt \cite{Puosi12SE} and experimental data for ortho-terphenyl (OTP) \cite{tolleRepProg01,Sillescu_JPCB97}.  {\it All} the numerical and the experimental data presented in Figure \ref{figSEUniv} exhibit the universal scaling expressed by Eq.\ref{eqn:parabola2}, see top inset for the BM and MA systems and ref.\cite{OurNatPhys,OttochianLepoJNCS11} for the polymer melt and OTP. Figure \ref{figSEUniv} is the major result of the present paper. It evidences the  scaling of the SE violation in terms of the DW in three different numerical atomic and polymeric models ( BM, MA, polymer melt) and OTP.  {Consideration of the data above $\langle u^2 \rangle/u^2_g \sim 10$ in terms of the vibrational scaling is not possible since cage effects are negligible  \cite{OurNatPhys}.  Alternative definition of the SE product as $K'_{SE}=D\eta/T$ (or $K'_{SE}=D\tau_\alpha/T$) virtually does not alter the quality of the scaling, as shown in the bottom inset of Fig.\ref{figSEUniv} for the BM and MA systems. 
 Notice that Fig.\ref{figSEUniv} presents results for polymers with {\it different} lengths since $K_{SE}/K_0$ is {\it independent} of it \cite{Puosi12SE}.
 
 We now perform a severe test of the vibrational scaling proposed in ref. \cite{OurNatPhys} by deriving an expression with {\it no} adjustable parameters of the master curve evidenced by Figure \ref{figSEUniv}. To this aim, we resort to the usual interpretation of the SE breakdown in terms of DHs, the spatial distribution of the characteristic relaxation times $\tau$ developing close to GT \cite{Ediger00,SillescuSEJNCS94,BerthieRev}. We are interested in the quantity $D \tau_\alpha$.
We define the macroscopic diffusivity as $D = \langle a^2/ 6 \tau_\alpha^{(HW)} \rangle_{a^2}$ and, as in the derivation of Eq.\ref{eqn:parabola2}, take $\tau_\alpha = \langle \tau_\alpha^{(HW)} \rangle_{a^2}$.
 The resulting expression of the quantity $D \tau_\alpha$ is a function of  $\langle u^2 \rangle/u^2_g$ with {\it no} adjustable parameters since it involves the universal parameters $\hat{\beta}$ and $\hat{\gamma}$ of Eq.\ref{eqn:parabola2}.
The corresponding ratio  $K_{SE}/K_0$ reads: 
\begin{eqnarray}\label{eqn:Dtau}
  && K_{SE}(x)/K_0=  \exp \left [  2 \hat{\gamma}/x^{2} \right ] \times  \nonumber \\  &&  \frac{ \left [ 1+\mbox{erf} \left ( \frac{\hat{\gamma}^{1/2}}{x} + \frac{\hat{\beta}}{2\hat{\gamma}^{1/2}}  \right ) \right] \left [ \mbox{erfc}  \left ( \frac{\hat{\gamma}^{1/2}}{x} - \frac{\hat{\beta}}{2\hat{\gamma}^{1/2}}  \right ) \right]}{\left [  1+ \mbox{erf} \left ( \frac{\hat{\beta}}{2\hat{\gamma}^{1/2}} \right) \right]^2} 
\end{eqnarray}
where $x= \langle u^2 \rangle/u^2_g$, $\mbox{erfc}(x) = 1- \mbox{erf}(x)$  and $\mbox{erf}(x)$ is the error function. 
   The result, shown in Fig.\ref{figSEUniv} (dark-blue curve), suggests that, even if the form of the distribution of the square displacements needed to overcome the relevant energy barriers $p(a^2)$ is adequate for {\it large} displacements governing $\tau_\alpha$\cite{OurNatPhys,lepoJCP09,Puosi11,SpecialIssueJCP13,UnivSoftMatter11}, it must be improved for {\it small} displacements affecting $D$. 
 Still, the exponential factor in Eq. \ref{eqn:Dtau}, controlling the SE breakdown, corresponds to the quadratic term in Eq. \ref{eqn:parabola2}, supporting the interpretation of the latter as due to dynamical heterogeneities \cite{OurNatPhys}. 
 Alternatively, we assume  the FSE form  $D \tau_\alpha \simeq \tau_\alpha ^{-\kappa}\tau_\alpha \simeq \tau_\alpha ^{1-\kappa}$ and $\tau_\alpha$ as given from Eq. \ref{eqn:parabola2} so that $K_{SE}/K_0 \simeq (\tau_\alpha/\tau_0)^{1-\kappa}$. Best-fit is found for $\kappa = 0.85$ (orange curve in Fig.\ref{figSEUniv}), which interestingly equals the universal value found by Mallamace et al \cite{MallamaceFSEPNAS10}.  
 
 The present results strongly suggest that the vibrational scaling in terms of the reduced DW factor $\langle u^2 \rangle/u^2_g$ encompasses the DH influence on the SE breakdown. A similar conclusion was reached by evaluating the DW factor of a simulated 2D glassformer in a time lapse being one order of magnitude longer than the one setting $\langle u^2 \rangle/u^2_g$ \cite{Harrowell06}. 
  Even if the experimental and the MD results are fairly scaled to a master curve by the reduced DW factor $\langle u^2 \rangle/u^2_g$, the proposed universal character of this scaling has to be corroborated by further investigations.  Two distinct guidelines are in order: i) a wider range of simulated and experimental systems, the latter at present time being limited mainly by the lack of DW data, ii) a better description of the universal  master curve with respect to the one provided by Eq.\ref{eqn:Dtau} by improving the form of the distribution of the squared displacements controlling  the structural relaxation and diffusion, in particular in the part that affects the latter.

\begin{acknowledgments}
F.P., A.P. and N.J.  acknowledge the financial support from the Centre of Excellence of Multifunctional Architectured Materials ``CEMAM'' No ANR-10-LABX-44-01 funded by the ``Investments for the Future'' Program. This work was granted access to the HPC resources of IDRIS under the allocation 2017-A0020910083 made by GENCI. Some of the computations presented in this paper were performed using the Froggy platform of the CIMENT infrastructure (https://ciment.ujf-grenoble.fr), which is supported by the Rh\^{o}ne-Alpes region and the Equip@Meso project (reference ANR-10-EQPX-29-01) of the programme Investissements d'Avenir supervised by the Agence Nationale pour la Recherche. DL and FP acknowledge a generous grant of computing time from IT Center, University of Pisa and Dell EMC${}^\circledR$  Italia. 

\end{acknowledgments}

\end{document}